\documentclass[11pt]{article}
\usepackage{amssymb}
\author{K.Anguige\\K.P.Tod}
\title{\textbf{Isotropic Cosmological Singularities II.\\The Einstein-Vlasov system.}}
\begin{document}
\maketitle
\begin{abstract}
We consider the conformal Einstein equations for massless
collisionless gas cosmologies which admit an isotropic
singularity. After developing the general theory, we restrict to 
spatially-homogeneous cosmologies. We show that the Cauchy problem for these equations is well-posed
with data consisting of the limiting particle distribution
 function at the singularity.
\end{abstract}

\section{Introduction}
In the accompanying paper, (Anguige and Tod 1998, hereafter ATI), we considered
the Cauchy 
problem for the conformal Einstein equations with a perfect fluid 
source and with data given at an isotropic singularity. Among other 
results, we were able to show that a perfect fluid cosmology, with a 
polytropic equation of state, for which the Weyl tensor vanished at 
an initial, isotropic singularity was conformally-flat: if the Weyl 
tensor is zero initially then it is always zero. This is a strong 
result, and one naturally wonders how robust it is. In particular, 
does it continue to hold with other matter models?\\

In this, the second paper of the series, we consider the same Cauchy 
problem but with a source given by a different matter model, a 
massless collisionless gas. This is the Einstein-Vlasov system. Our 
long-term aim is to prove existence and uniqueness for this system 
with data at an isotropic singularity. However, in the present paper 
we only partially achieve this aim: we are able to prove what we want 
only with the restriction of spatial homogeneity. This is sufficient 
to investigate the Weyl tensor problem described above, but of course 
the long-term aim remains.\\

In the Einstein-Vlasov system (see e.g. Ehlers 1971, Rendall 1992, 1997) 
the matter is determined by the 
distribution function $f(x^a,p_b)$, a positive function on the 
cotangent bundle of space-time, which is interpreted as the density 
of particles at the space-time point labelled by $x^a$ which have 4-momentum $p_b$. 
The collisionless assumption is translated as the assumption that the 
distribution function is constant along the geodesic flow, giving an 
equation known in this context as the Vlasov (or Liouville) equation. The 
Einstein equations are written down with a stress-tensor defined as 
an invariant second moment of $f$ integrated over the fibre of the 
cotangent bundle. Thus $f$ is constrained by the condition that this 
moment exist. We will consider only massless matter, which means that 
$f$ is supported on the null-cone at each point. For analytical 
convenience, we will assume that $f$ is supported compactly on the 
cone and away from the vertex of the cone.\\

The massless assumption will imply that the theory has a good
transformation  
under conformal rescaling of the metric. For example, the 
distribution function is constant along null geodesics, which are 
conformally-invariant. Also the stress-tensor is trace-free, which is 
the case for radiation perfect-fluid where it is known from ATI 
that the conformal
factor can be chosen to be smooth at an isotropic singularity.\\

We refer to the accompanying paper for the background to conformal rescaling. 
Recall only that the physical metric $\tilde{g}_{ab}$ of space-time 
$\tilde{M}$ is related by rescaling to an unphysical metric $g_{ab}$ on a larger space-time $M$ according to:
$$ \tilde{g}_{ab} = \Omega^2g_{ab} $$
where the conformal factor $\Omega$ vanishes at a smooth surface 
$\Sigma$ in the extended space-time $M$. As remarked above, the 
massless assumption will 
enable us to suppose that $\Omega$ is smooth at $\Sigma$.\\

After establishing the conformal transformation properties of the 
matter model, we shall seek Bondi-type expansions of space-time 
quantities as power-series in the conformal factor, which becomes a 
good time-coordinate in the unphysical (rescaled) space-time. In this 
way, we are able to isolate the degrees of freedom of the problem, or 
equivalently to identify the Cauchy data. The data are very different 
from the perfect-fluid case. We need the first and second fundamental 
forms of the singularity surface $\Sigma$ and the initial 
distribution function $f^0$. These turn out to have very strong 
constraints among them, so strong in fact that {\em the initial 
distribution function $f^0$ uniquely determines the first and second 
fundamental forms of the singularity surface}. The initial 
distribution function is subject to a single, integral constraint 
which can be interpreted as a `zero drift' condition.\\

Recall from ATI that in the perfect-fluid case, the second 
fundamental form of $\Sigma$ is necessarily zero and the first 
fundamental form is free data. There is no free data for the matter 
(apart from the equation of state). One can say aphoristically that 
{\em the metric determines the matter}. From the previous paragraph, 
we see that circumstances are quite different in the Einstein-Vlasov 
case where {\em the matter determines the metric}.\\

Because the second fundamental form of the singularity necessarily 
vanishes for perfect fluid, the condition of initially-vanishing Weyl 
tensor forces the physical cosmology to be FRW, so that, as we noted 
above, the Weyl tensor is always zero if it is initially zero 
(ATI). With Einstein-Vlasov the situation is rather different. The 
second fundamental form is determined by $f^0$ and its vanishing is 
an integral constraint on $f^0$, related to the vanishing of a third 
moment. It is quite possible, as we shall see in section 6.6, that the 
lower moments of $f^0$ are such that the initial Weyl tensor is zero 
but there are higher moments which allow the later evolution to 
diverge from FRW, so that non-zero Weyl tensor emerges.\\

Once we have identified the data, we turn to the problem of 
existence and uniqueness of solutions. For this, we restrict to spatially-
homogeneous cosmologies. Then the Einstein equations become a system 
of ordinary differential equations with a singularity in the `time' 
and a source involving integrals over the distribution function $f$, 
while the Liouville equation remains a partial differential equation 
for $f$. We prove existence and uniqueness for this coupled 
system by a method based on results of Rendall (1994) and Rendall 
and Schmidt(1991).\\

The contents of the paper are as follows. We begin in section 2 with 
a review of relativistic kinetic theory and then in section 3 
consider the problem of conformally-rescaling the massless
Einstein-Vlasov 
(henceforth EV) system. The Bondi expansions are begun in 
section 4, which includes the proof that the constraint relating 
$f^0$ and the metric $a_{ij}$ of $\Sigma$ determines $a_{ij}$ 
uniquely. In section 5, we fix the conformal gauge for the EV system
which enables us to simplify the expressions found for the data in
section 4 by eliminating some gauge quantities. Section 6 is the heart
of the paper. We adapt the expansions of section 4 to a 
spatially-homogeneous cosmology, identify the data and then
prove, in Theorem 6.1, that solutions exist and are unique given
suitable data. We consider what data lead to FRW cosmologies and show
in particular that there are Bianchi-type V solutions in which the
Weyl tensor is zero initially but 
not for all time. \\

\section{Relativistic kinetic theory}
In kinetic theory, the matter content of spacetime~$\tilde{M}$~is
taken to consist of a
collection of particles which move on geodesics between
collisions. The motion of a 
freely falling particle is determined by the Lagrangian:
\begin{equation}L(x^{c},\dot{x}^{d})=\frac{1}{2}g_{ab}\dot{x}^{a}\dot{x}^{b}\end{equation}
since the Euler-Lagrange equations are equivalent to the (affine) geodesic
equation.
Writing~$p_{a}=g_{ab}\dot{x}^{b}=\frac{\partial L}{\partial\dot{x}^{a}}=$~canonical
momentum, we get the Hamiltonian
\begin{equation}H(x^{c},p_{d})=\frac{1}{2}g^{ab}p_{a}p_{b}\end{equation}
and thus we obtain canonical equations for the geodesics. ~$H$~is
constant along each particle world-line,~$p_{a}$~is the
4-momentum, and~$m=\sqrt{p_{a}p^{a}}$~is the mass of the test
particle. To take into account all possible particle positions and
momenta, we must consider the one-particle phase space~$P$~. This is
the collection of instantaneous states~$(x^{a},g_{ab}\dot{x}^{b})$~of a
single particle and is the future causal part of the spacetime cotangent
bundle.\\

States having a given mass~$m\geq 0$~form the seven-dimensional
submanifold $P_{m}$ of $P$. On ~$P_{m}$~we take~$(x^{a},p_{i})$~as local
coordinates, ~$p_{0}$~being determined by the
equation~$g^{ab}p_{a}p_{b}=m^{2}$, and the requirement that~$p_{a}$~be
future directed (conventionally, $a,b,c,..=0-3$ while
$i,j,k,..=1-3$). The free-fall trajectories define on~$P$~a congruence
of curves, along which~$H$~is constant. The geodesic spray~$\mathcal{L}$~is a vector
field along these curves defined, in local coordinates, by
\begin{equation}\mathcal{L}=g^{ab}p_{a}\frac{\partial}{\partial
x^{b}}-\frac{1}{2}p_{a}p_{b}\frac{\partial g^{ab}}{\partial
x^{c}}\frac{\partial}{\partial p_{c}}\end{equation}

The cotangent space to~$M$~at~$x$~is a flat Lorentzian manifold, and on
the submanifold~$P_{m}(x)$~there exists an invariant volume measure~$\pi_{m}$~ given by
\begin{equation}\pi_{m}=\frac{(-g)_{x}^{-1/2}}{p^{0}}d^{3}p_{i}.\end{equation}
The distribution of particles and momenta is described by a scalar
function~$f=f(x^{a},p_{b})$~on~$P$, and the central result of
relativistic kinetic theory is the equation
\begin{equation}\mathcal{L}(f)=g\end{equation}
where~$g$~describes the density of particle collisions. Hence the condition
that~$f$~represents a collisionless gas is simply
\begin{equation}\mathcal{L}(f)=0\end{equation}
and this is known as the Vlasov equation. Note that~$f$~satisfies the
Vlasov equation if and only if it is constant along geodesics.\\

The stress-energy-momentum tensor due to particles of mass~$m$~is
given by
\begin{equation}T^{m}_{ab}(x)=\int_{P_{m}(x)}fp_{a}p_{b}\pi_{m}\end{equation}
and if the Vlasov equation is satisfied then
\begin{equation}\nabla^{a}T^{m}_{ab}=0\end{equation}
If the gravitational field is due solely to particles of a single
mass~$m$, represented
by~$f$, then the coupled Einstein-Vlasov equations, for the
metric~$g_{ab}$~and the particle distribution~$f$~are
\begin{equation}G_{ab}=8\pi\int_{P_{m}}fp_{a}p_{b}\pi_{m}\end{equation}
\begin{equation}\mathcal{L}_{g}(f)=0\end{equation}

\section{Conformally rescaling the massless EV equations}
Now we shall restrict to massless, collisionless particles and
investigate the transformation properties of the Einstein-Vlasov system
under conformal rescaling. For a future-pointing, null co-vector $p_a$, the component $p_0$, either coordinate component or orthonormal-frame component, is determined by the spatial components $p_i$. Thus we write ~$\tilde{f}(x^{\alpha},p_{i})$~for the distribution function for
massless particles on~$P_{0}$~in a
spacetime~$(\tilde{M},\tilde{g}_{ab})$. Suppose that
~$\tilde{g}_{ab}$~and~$\tilde{f}$~satisfy tilded versions of
(9)-(10). Let~$g_{ab}$~be the metric defined by
\begin{equation}\tilde{g}_{ab}=~\Omega^{2}g_{ab}\end{equation}
Under this transformation, null geodesics of~$\tilde{g}_{ab}$~become
null geodesics of~$g_{ab}$. Thus, since~$\tilde{f}$~is constant along the
geodesic flow
in~$(\tilde{M},\tilde{g}_{ab})$,~$f\equiv\tilde{f}$~will be constant
along the geodesic flow in~$(M,g_{ab})$. In fact one calculates that
\begin{equation}
\mathcal{L}_{\tilde{g}}\tilde{f}=\Omega^{-2}\mathcal{L}_{g}f
\end{equation}
and it follows that~$f$~satisfies the Vlasov equation in~$M$.
\\If we now define
\begin{equation}T_{ab}=\int~fp_{a}p_{b}\frac{(-g)_{x}^{-1/2}}{p^{0}}d^{3}p\end{equation}
then the conservation equation~$\nabla^{a}T_{ab}=0$~holds in~$M$, just as the tilded version holds 
in the physical spacetime. This
follows either from the unphysical Vlasov equation, or the relation
\begin{equation}\tilde{T}_{ab}=\frac{1}{\Omega^{2}}T_{ab}\end{equation}
together with~$\tilde{T}^{a}_{a}=0$, which in turn follows from~$m=0$.\\

With the conformal transformation of the Ricci tensor taken
e.g. from ATI, we can now write down the conformal EV 
equations for~$g_{ab}$~and~$f$~as
\begin{displaymath}R_{ab}=2\nabla_{a}\nabla_{b}\log\Omega-2\nabla_{a}\log\Omega\nabla_{b}\log\Omega\end{displaymath}
\begin{equation}+g_{ab}(\square
\log\Omega+2\nabla_{c}\log\Omega\nabla^{c}\log\Omega)+\frac{8\pi}{\Omega^{2}}\int~fp_{a}p_{b}\frac{(-g)^{-1/2}}{p^{0}}d^{3}p\end{equation}
and
\begin{equation}\mathcal{L}_{g}(f)=0\end{equation}
Note that (15) implies
\begin{equation}\square\Omega=\frac{1}{6}R\Omega\end{equation}
since~$T_{a}^{~a}=0$.\\

Now, as~$\Omega$~is smooth in~$\tilde{M}$~and satisfies the regular, linear
wave equation (17) in~$M$, one must have that~$\Omega$~has a smooth
extension onto~$M$~(see Racke 1992). It follows that~$\Omega$~is a
smooth function in~$M$.\\

In the sequel it will be important that~$\Omega$~be a good
coordinate in~$M$, and thus for we make the following assumption:
\\\\\textbf{Assumption~3.1}~\textit{The conformal
factor}~$\Omega$~
\textit{is such that}~$\nabla_{a}\Omega\neq
0$~\textit{at}~$\Sigma$. \\\\Thus~$\Omega$~corresponds to the time
function~$Z$~of ATI in the case of a radiation fluid, and for
convenience we will henceforth write~$\Omega=Z$.
\section{The initial data: Bondi expansions near~$\Sigma$.}
The singular nature of the field equation (15) imposes constraints
on the data at the singularity~$\Sigma$~. To see what these
constraints are we shall seek `Bondi' expansions of the field
variables in powers of~$Z$~near~$\Sigma$ (analogous to the
corresponding expansions near future-null-infinity, whence the name). That
is, we write
\begin{displaymath}h_{ij}=h^{0}_{ij}+Z
h^{1}_{ij}+\ldots\end{displaymath}
etc, and compare coefficients in (15)-(16).
To aid calculation we use coordinates~$x^a$~,
with~$x^{0}=Z$~and the ~$x^{i},~(i=1,2,3)$~taken to be comoving along the integral
curves of ~$\nabla^{a}Z$. In these coordinates the line element
in~$M$~can be written
\begin{equation}ds^{2}=\frac{1}{V^{2}}dZ^{2}-h_{ij}dx^{i}dx^{j}\end{equation}
with~$h_{ij}$~positive definite and $V^2 = g^{ab}Z_{,a}Z_{,b}$.\\

The Vlasov equation for ~$f=f(x^a,p_{i})$~takes the form
\begin{equation}V^{2}p_{0}\frac{\partial
f}{\partial Z}-h^{ij}p_{i}\frac{\partial f}{\partial
x^{j}}-\frac{1}{2}\{(\partial_{i}V^{2})(p_{0})^{2}-(\partial_{i}h^{mn})p_{m}p_{n}\}\frac{\partial
f}{\partial p_{i}}=0\end{equation}
where~$V^{2}(p_{0})^{2}=h^{ij}p_{i}p_{j}$.\\

The Einstein equations are just
\begin{equation}Z^{2}R_{ab}-2Z\nabla_{a}\nabla_{b}Z+4\nabla_{a}Z\nabla_{b}Z=g_{ab}(Z\square
 Z+V^{2})+8\pi
T_{ab}\end{equation}
\subsection{O(1) terms}
Considering the lowest order terms in (20) gives
\begin{equation}8\pi(T_{00})^0=3,~~~8\pi(T_{0i})^0=0,~~~8\pi(T_{ij})^0=(V^{0})^{2}h^{0}_{ij}.\end{equation}
The second of these is just an integral constraint on~$f^{0}$, the
distribution function at $\Sigma$:
\begin{equation}\int f^{0}(x^j,p_k)p_{i}~d^{3}p=0;\end{equation}
the third gives an implicit relation between~$f^{0}$~and~$h^{0}_{ij}$:
\begin{equation}(V^{0})^{2}h^{0}_{ij}=\frac{8\pi}{\sqrt{h^{0}}}\int\frac{f^{0}p_{i}p_{j}}{(h^{mn}p_{m}p_{n})^{1/2}}~d^{3}p\end{equation}
where~$h^{0}=\textrm{det}(h^{0}_{ij})$~and the first is implied by 
the third since~$T_{ab}$~is trace-free.\\

\noindent It is now natural to ask two questions:
\begin{enumerate}
\item Given a positive, suitably integrable function~$f^0$, does
there exist a 3-metric~$h^0_{ij}$~satisfying equation (23)?
\item Given a positive-definite 3-metric~$h^0_{ij}$~does there exist a
positive function~$f^0$~satisfying equation (23)?
\end{enumerate}
These questions are answered affirmatively by Theorems 4.1 and 4.2 below.
\\\\
\textbf{Theorem~4.1}~Let~$f(x^{j},p_{i})$~be a smooth, positive
function on~$U\times\mathbb{R}^{3}$, where~$U$~is an open subset of~$\mathbb{R}^{3}$. Suppose that for each~$x\in U$
\begin{enumerate}
\item $f$~is compactly supported in~$p$;
\item $f$~is supported outside some open ball
containing~$p=0$;
\item $f$~is not identically zero in~$p$.
\end{enumerate}
Then, given a smooth strictly positive function~$V(x)$~on~$U$, there exists a unique positive definite 3-metric~$h_{ij}(x)$~on$~U$
~satisfying (23). Moreover, the metric~$h_{ij}(x)$~is smooth.\\

\textit{Proof.}~The metric is found by a minimisation property. First define the function~$F$~by
\begin{equation}F(b^{11},b^{22},b^{33},b^{12},b^{13},b^{23})=(\textrm{det}B)^{-1/6}\int
(p^{T}Bp)^{1/2}f~d^{3}p\end{equation}
where
\begin{equation}B=\left(\begin{array}{ccc}b^{11}&b^{12}&b^{13}\\b^{12}&b^{22}&b^{23}\\b^{13}&b^{23}&b^{33}\end{array}\right)\end{equation}
and~$B$~is positive definite.\\

Clearly~$F(B)=F(\lambda B)$~for~$\lambda\in\mathbb{R}$, so we can think of~$F$~as a function
defined on a region~$S$~in~$\mathbb{R}^6$~consisting of
points~$P=(b^{11},b^{22},b^{33},b^{12},b^{13},b^{23})$~for which~$B$~is
positive definite and~$\textrm{tr}B=1$.\\

Now for a couple of Lemmata:
\\\\\textbf{Lemma~4.1.}~The function~$F$~has a critical point:
\begin{equation}\frac{\partial F}{\partial b^{ij}}=0~~~~i=1,2,3~~j\geq
i\end{equation}
if and only if the following equation holds
\begin{equation}a_{ij}\int
(p^{T}Bp)^{1/2}f~d^{3}p=3\int\frac{fp_{i}p_{j}}{(p^{T}Bp)^{1/2}}~d^{3}p\end{equation}
where~$(a_{ij})=(b^{ij})^{-1}$.\\

\textit{Proof}.~Note that
\begin{equation}\frac{\partial (\textrm{det}B)}{\partial
b^{ij}}=\frac{2\hat{b}_{ij}}{2^{(\delta_{ij})}}\end{equation}
where~$\hat{b}_{ij}=(\textrm{det}B)a_{ij}$, and there is no summation
on the rhs.\\

Hence
\begin{displaymath}\frac{\partial F}{\partial
b^{ij}}=-\frac{1}{6}(\textrm{det}B)^{-1/6}\frac{2}{2^{(\delta_{ij})}}a_{ij}\int
(p^{T}Bp)^{1/2}f~d^{3}p\end{displaymath}
\begin{equation}+(\textrm{det}B)^{-1/6}\frac{1}{2^{(\delta_{ij})}}\int
(p^{T}Bp)^{-1/2}p_{i}p_{j}f~d^{3}p\end{equation}
and the lemma follows.
\\\\
\textbf{Lemma~4.2}~If~$M=(m^{ij})$~is a~$3\times 3$~matrix
with~$\textrm{tr}M=0$, then
\begin{equation}\delta^{2}F\equiv \frac{\partial^{2}F}{\partial
b^{ij}\partial
b^{kl}}m^{ij}m^{kl}2^{(\delta_{kl}-1)}2^{(\delta_{ij}-1)}>0\end{equation}
at any critical point of~$F$.\\

\textit{Proof}.~By definition one has
\begin{equation}\hat{b}_{ij}=\frac{1}{2}\epsilon_{ipq}\epsilon_{jrs}b^{pr}b^{qs}\end{equation}
which implies
\begin{equation}\frac{\partial\hat{b}_{ij}}{\partial
b^{kl}}=\sum_{p,r}~\frac{1}{2^{(\delta_{kl})}}(\epsilon_{ipk}\epsilon_{jrl}+\epsilon_{ipl}\epsilon_{jrk})b^{pr}\end{equation}
and thus
\begin{displaymath}\frac{\partial^{2}F}{\partial b^{ij}\partial
b^{kl}}=2(1-\delta^{ij})\Bigg\{-\frac{1}{6}\hat{b}_{ij}(\textrm{det}B)\frac{\partial
F}{\partial
b^{kl}}-\frac{1}{4}(\textrm{det}B)^{-1/6}2^{(1-\delta_{kl})}\int\frac{fp_{i}p_{j}p_{k}p_{l}}{(p^{T}Bp)^{3/2}}~d^{3}p\end{displaymath}
\begin{displaymath}-\frac{1}{12}(\textrm{det}B)^{-7/6}2^{(1-\delta_{kl})}\hat{b}_{kl}\int\frac{fp_{i}p_{j}}{(p^{T}Bp)^{1/2}}~d^{3}p\end{displaymath}
\begin{equation}-\frac{1}{6}F(B)\left(-((\textrm{det}B)^{-2})2^{(1-\delta_{kl})}\hat{b}_{kl}\hat{b}_{ij}+(\textrm{det}B)^{-1}\frac{1}{2^{(\delta_{kl})}}(\epsilon_{ipk}\epsilon_{jrl}+\epsilon_{ipl}\epsilon_{jrk})(b^{pr})\right)\Bigg\}\end{equation}
At a critical point we may, without loss of generality, take~$B=I_{3}$~, the $3\times 3$~identity matrix, 
for if~$b^{ij}$~solves
(27) then~$I_{3}$~solves the same equation with ~$f$~replaced
by~$\hat{f}$, where~$\hat{f}(x,p)=f(x,Lp)$~for some linear
transformation~$L$, and we can work with~$\hat{f}$.\\

Now drop the hat and evaluate (33) at a critical point to give
\begin{displaymath}\frac{\partial^{2}F}{\partial
b^{ij}b^{kl}}=2^{(1-\delta_{ij})}2^{(1-\delta_{kl})}\Bigg\{-\frac{1}{4}\int\frac{fp_{i}p_{j}p_{k}p_{l}}{(p^{T}{p})^{3/2}}~d^{3}p-\frac{1}{12}\delta_{kl}\int\frac{fp_{i}p_{j}}{(p^{T}p)^{1/2}}~d^{3}p\end{displaymath}
\begin{equation}-\frac{1}{6}F(I_{3})\Big(-\delta_{kl}\delta_{ij}+\frac{1}{2}(\epsilon_{ipk}\epsilon_{jrl}+\epsilon_{ipl}\epsilon_{jrk})\delta^{pr}\Big)\Bigg\}\end{equation}
so that
\begin{displaymath}\frac{\partial^{2}F}{\partial b^{ij}\partial
b^{kl}}m^{ij}m^{kl}2^{(\delta_{kl}-1)}2^{(\delta_{ij}-1)}=-\frac{1}{4}\int\frac{(p^{T}Mp)^{2}f}{(p^{T}p)^{3/2}}~d^{3}p-\frac{1}{12}\textrm{tr}M\int\frac{(p^{T}Mp)f}{(p^{T}p)^{1/2}}~d^{3}p\end{displaymath}
\begin{equation}-\frac{1}{6}F(I_{3})(-(\textrm{tr}M)^{2}+2\textrm{tr}\hat{M})\end{equation}
Note that for any~$3\times 3$~matrix~$M$~there is the following relation:
\begin{equation}\textrm{tr}\hat{M}=\frac{1}{2}\{(\textrm{tr}M)^{2}-\textrm{tr}M^{2}\}\end{equation}
where~$\hat{M}$~is the matrix of cofactors.
Hence (35) becomes
\begin{equation}\delta^{2}F=-\frac{1}{4}\int\frac{(p^{T}Mp)^{2}f}{(p^{T}p)^{3/2}}~d^{3}p~+\frac{1}{6}F(I_{3})\textrm{tr}M^{2}\end{equation}
Now let the eigenvalues
of~$M$~be~$\lambda_{1},\lambda_{2},\lambda_{3}$~with\linebreak
~$\lambda_{1}\geq\lambda_{2}>0$,
and~$\lambda_{3}=-(\lambda_{1}+\lambda_{2})$~(replacing~$M$~by~$-M$~if
necessary).\\
It follows that
\begin{equation}|p^{T}Mp|^{2}\leq |\lambda_{3}|^{2}|p^{T}p|^{2}\end{equation}
and equality holds for only one value of~$p$ (up to scale).
\\Hence
\begin{equation}\delta^{2}F>\frac{1}{12}F(I_{3})(\lambda_{1}-\lambda_{2})^{2}\geq
0\end{equation}
Note that we have actually shown that~$\hat{F}$, defined by
\begin{displaymath}\hat{F}(B)=(\textrm{det}B)^{-1/6}\int(p^{T}Bp)^{1/2}\hat{f}~d^{3}p\end{displaymath}
is such that~$\delta^{2}\hat{F}>0$~at~$B=I_{3}$.~But we have
\begin{displaymath}F(B)\equiv|\textrm{det}L|^{-4/3}\hat{F}((L^{T})^{-1}BL^{-1})\end{displaymath}
It follows that~$\delta^{2}F>0$~at a critical point~$B$. Hence the lemma.
\\\\
Suppose next that (27) holds and put~$h_{ij}=\lambda a_{ij}$~where
\begin{displaymath}\lambda^{2}=\left(\frac{8\pi}{3V^{2}(\textrm{det}A)^{1/2}}\right)\int(p^{T}Bp)^{1/2}f~d^{3}p\end{displaymath}
Then ~$h_{ij}$~solves (23) and so there is a one-one relation
between critical points of~$F$~and solutions of (23).
\\
Returning to the set~$S$~on which~$F$~is defined, we see easily that
it is both convex and open
in~$O\equiv\mathbb{R}^{6}\cap\{B:~\textrm{tr}B=1\}$. Since~$B$~is +ve
definite we get from (36)
that~$(\textrm{tr}B)^{2}\geq\textrm{tr}B^{2}$~and
consequently~$\textrm{max}\{|b^{12}|,|b^{13}|,|b^{23}|\}\leq\frac{1}{2}$,
so that~$S$~is bounded.
\\
Since~$S$~is convex it must just consist of radii from some point
in~$S$, in the directions spanned by the~$b^{12},b^{13},b^{23}$~axes, and in the plane
given by
\begin{displaymath}b^{11}+b^{22}+b^{33}=1\end{displaymath}
 The convexity also implies
that~$S$~has continuous radius, and thus is homeomorphic to an open
ball in~$\mathbb{R}^{5}$. In particular~$S$~is contractible: it has
the homotopy type of a point. By Lemma 3.2~$\delta^{2}F>0$~at a
critical point, when variations are restricted to directions in~$S$,
so that any critical point of~$F$~is a minimum.
\\
Now~$F\rightarrow\infty$~on the boundary~$\partial S$~of~$S$~(thinking
of~$S$~as a subset of~$O$), so
that~$\frac{1}{F}$~is continuous on~$\bar{S}$, with~$\frac{1}{F}=0$~on~$\partial
S$. Now choose~$a>0$~and consider the following subset~$Q$~of$~S$:
\begin{displaymath}Q=\Big\{s\in
S:\frac{1}{F}\geq\frac{1}{a}\Big\}\end{displaymath}
Since~$S$~is bounded and~$\frac{1}{F}$~is continuous, one must have
that~$Q$~is a closed, bounded subset of~$O$, disjoint from~$\partial S$. Writing~$G=-F$~ and~$b=-a$~now gives
that~$G^{-1}[b,\infty)$~is compact. Since~the critical points
of~$G$~are all maxima, it follows from theorem 3.5 of (Milnor 1963) 
that~$S$~has the homotopy type of a discrete set of points, one for
each critical point of ~$G$. But we already know
that~$S$~has the homotopy type of a single point and thus~$F$~has just one critical point, as required.
\\\\
It remains to show that the metric determined by (23) is smooth. To
do this first write
\begin{equation}H_{ij}\equiv \frac{\partial F}{\partial
b^{ij}}\end{equation}
and suppose that~$B^{0}$~is the critical point of~$F$~at~$x^{0}$,~so
that~$H_{ij}(B^{0},x^{0})=0$.\\
Now, by hypothesis,~$H_{ij}$~is smooth in~$B$~and~$x$~in a
neighbourhood of~$(B^{0},x^{0})$, and we have
\begin{equation}\frac{\partial H_{ij}}{\partial
b^{kl}}=\frac{\partial^{2}F}{\partial b^{kl}\partial
b^{ij}}\end{equation}
But from Lemma 4.2 we have that~$\delta^{2}F>0$~at~$(B^{0},x^{0})$,
and thus~$(d_{B}H)$~is invertible at this point. From the inverse
function theorem it now follows that~$B(x)$~defined by~$H_{ij}(B,x)=0$~depends smoothly
on~$x$.
\\\\
\textbf{Theorem~4.2}~Let~$h_{ij}(x)$~be a smooth, positive-definite
metric on an open set~$U\subset\mathbb{R}^{3}$, with~$V$~a smooth,
strictly positive function on~$U$. Then there exists a
smooth positive function~$f$~on~$U\times\mathbb{R}^{3}$~satisfying
(23), and in fact there are many such~$f$~for each~$h_{ij}$.

\textit{Proof}.~First note that equation (23) can be written in matrix form as
\begin{equation}(V^{2}\sqrt{\textrm{det}H}/8\pi)H(x)=\int\frac{f(x,p)pp^{T}}{(p^{T}H^{-1}p)^{1/2}}~d^{3}p\end{equation}
where~$H=(h_{ij})$.\\
Now let~$\phi :\mathbb{R}\rightarrow\mathbb{R}$~be a smooth
function and~$r_{1},~r_{2}$~positive constants, chosen so that
\begin{enumerate}
\item $\phi (x)=0$~for~$|x|\leq r_{1}$
\item $\phi (x)=0$~for~$|x|\geq r_{2}$
\item $\phi (x)>0$~for~$r_{1}<|x|<r_{2}$
\item
\end{enumerate}
\begin{displaymath}I_{3}=\int\frac{\phi(p^{T}p)}{(p^{T}p)^{1/2}}pp^{T}~d^{3}p\end{displaymath}
If we let~$f(x,p)=V^{2}(x)\phi(p^{T}Hp)$, then~$f$~is smooth and
solves (42). Clearly there is considerable freedom in the choice of~$\phi$.

\subsection{O(Z)~terms}
Let~$K_{ab}$~be the second fundamental form in ~$M$, so
that~$K_{ij}=-\frac{1}{2}V\partial_{z}h_{ij}$. A calculation of the O(Z)
terms in the space-space components of (20) gives
\begin{equation}\Big(2\delta^{i}_{~k}\delta^{j}_{~l}-\chi^{ij}_{~~kl}\Big)(K^{0})^{kl}=D^{m}(\chi^{ij}_{~~m})+\frac{1}{2}(K^{0})(h^{0})^{ij}\end{equation}
where~$K^{0}=(h^{0})^{ij}K^{0}_{ij}$
\begin{equation}\chi_{ijkl}\equiv\frac{4\pi}{(V^{0})^{2}\sqrt{h^{0}}}\int\frac{f^{0}p_{i}p_{j}p_{k}p_{l}}{((h^{0})^{mn}p_{m}p_{n})^{3/2}}~d^{3}p\end{equation}
\begin{equation}\chi_{ijk}\equiv
-\frac{4\pi}{(V^{0})^{2}\sqrt{h^{0}}}\int\frac{f^{0}p_{i}p_{j}p_{k}}{((h^{0})^{mn}p_{m}p_{n})}~d^{3}p\end{equation}
and~$D_{i}$~is the covariant derivative operator associated with~$(h^{0})_{ij}$.\\
Clearly the eigenvalues of~$\chi_{ijkl}$~regarded as a linear
transformation on $3 \times 3$ symmetric matrices are positive. Also,
from (23), the trace of this linear transformation is $3/2$, so that
each eigen-value is no greater than $3/2$. Thus the `matrix'
 acting on~$(K^{0})_{ij}$~in (43) is invertible, so that the
trace-free part of~$(K^{0})_{ij}$~is determined by~$f^{0}$~through
this equation. The trace is not determined but we shall see in 
section 5 that this can be removed by fixing the conformal gauge.\\

The O(Z) terms in the space-time and time-time components of (20)
impose no more constraints on the data at~$Z=0$.
      
\section{Conformal gauge fixing in EV}
Inherent in the definition of an isotropic singularity is the freedom
to choose a new conformal factor~$\bar{\Omega}$~and conformal
metric~$\bar{g}_{ab}$~by rescaling with a regular non-zero
function~$\Theta$~according
to~$\Omega\rightarrow\bar{\Omega}=\Theta\Omega$. 
We now use this freedom to make a 
conformal gauge choice which should be useful for solving the conformal EV equations in full generality.
\\\\
First recall the definition of~$V$~as~$V^{2}=g^{ab}\nabla_{a}Z\nabla_{b}Z$, and suppose one has
\begin{equation}\tilde{g}_{ab}=Z^{2}g_{ab}=\bar{Z}^{2}\bar{g}_{ab}
\end{equation}
Write~$h=\bar{Z}/Z$. Then at~$\Sigma$~one has
\begin{equation}\bar{V}^{2}=h^{4}V^{2}\end{equation}
Since~$\nabla_{a}Z\neq 0$~at~$\Sigma$, it follows that~$h$~can be chosen so that~$\bar{V}=1$~there.\\

Now write~$\hat{g}_{ab}=h^{2}\bar{g}_{ab}$,~for some new function~$h$,
with~$h=1$~at~$\Sigma$. If ~$\bar{K}$~is the trace of the second
fundamental form of~$\bar{g}_{ab}$, with~$\hat{K}$~defined similarly,
then at~$\Sigma$~one gets
\begin{equation}\hat{K}=\frac{1}{h}(\bar{K}+3\bar{N}^{d}\nabla_{d}\log
h)\end{equation}
where~$\bar{N}^{a}$~is the unit normal to~$\Sigma$~with respect
to~$\bar{g}_{ab}$. We can get $\hat{K}=0$~at~$\Sigma$~by
choosing~$h$~so that
\begin{equation}\bar{N}^{d}\nabla_{d}h=-\frac{1}{3}\bar{K}\end{equation}
there.\\

Consider now the following wave equation for~$h$:
\begin{equation}\bar{\square}h=\frac{1}{6}\bar{R}h\end{equation}
with initial
data~$h=1,~\bar{N}^{d}\nabla_{d}h=-\frac{1}{3}\bar{K}$.\\

It is standard (Racke 1992) that this equation has a unique smooth
solution~$h$~with the given data. If we
let~$\hat{Z}=\bar{Z}/h$,~then~$\tilde{g}_{ab}=\hat{Z}^{2}\hat{g}_{ab}$,
and from (17) there follows
\begin{equation}\widehat{\square}\widehat{Z}\equiv 0\end{equation}
We can now prove the following:
\\\\
\textbf{Lemma 5.1} If~$(\tilde{M},\tilde{g}_{ab})$~is a solution of the
massless Einstein-Vlasov equations, with an isotropic singularity, and
Assumption 3.1 holds, then the conformal factor~$Z$~may be chosen so
that~$V(0)=1,~K(0)=0$, and~$Z$~is a harmonic function
in~$M$. Moreover, these three choices fix the conformal factor completely.
 
\textit{Proof}.~Suppose~$\tilde{g}_{ab}=Z^{2}g_{ab}=\bar{Z}^{2}\bar{g}_{ab}$~,
with~$Z,~\bar{Z}$~chosen as above so that
\begin{displaymath}V(0)=\bar{V}(0)=1~,~~K(0)=\bar{K}(0)=0,~~\square Z=\bar{\square}\bar{Z}\equiv
0\end{displaymath}
Define ~$h$~by~$h=\bar{Z}/Z$, so that~$h(0)=1$~and~$(N^{d}\nabla_{d}h)(0)=0$. One has
\begin{equation}\bar{R}=h^{-2}(R-\frac{6}{h}(\square  h))\end{equation}
But from (17) ~$R=\bar{R}=0$, and thus
\begin{equation}\square h=0\end{equation}
Hence~$h\equiv 1$, and~$\bar{Z}=Z$.
\\\\

The initial data set for the conformal EV equations at~$Z=0$, in the gauge of Lemma 5.1, can now be summarised as follows:
\\
\begin{itemize}
\item The initial distribution function~$f^{0}$~is free data, subject only
to the integral constraint (22).
\item The initial
3-metric~$h^{0}_{ij}$~is determined by ~$f^{0}$~via~(23) \linebreak with ~$V^0\equiv 1$.
\item The initial second fundamental form~$K^{0}_{ij}$~is determined by
~$f^{0}$~via (43) with~$V^0\equiv 1$,~$\textrm{tr}K^{0}\equiv 0$
\end{itemize}
\section{The Cauchy problem}
We have seen that the Einstein-Vlasov system has good behaviour under 
conformal rescaling and that it is possible to specify Cauchy data for
the rescaled equations at an isotropic singularity. In this section 
we show that if attention is restricted to spacetimes with Bianchi 
type spatial symmetry, then this Cauchy data determines a unique 
solution of the field equations.

In this section only we will take the spacetime metric~$\tilde{g}_{ab}$~to be of signature ~$(-+++)$.

\subsection{Bianchi type spacetimes}
The Bianchi type spacetimes are a class of spatially homogeneous
cosmological models having a 3-parameter Lie group~$G$~of
isometries, transitive on space-like 3-surfaces, the surfaces of
homogeneity. They are classified into nine types according to the
nature of the Lie algebra associated with~$G$~(see Wald 1984 for
details). In such spacetimes there exists a cosmic time 
function~$t$~and one-form fields~$(e^{i})_{a}$~such that the metric 
can be written as
\begin{equation}\tilde{g}_{ab}=-\nabla_{a}t\nabla_{b}t+\tilde{a}_{ij}(t)(e^{i})_{a}(e^{j})_{b}\end{equation}
The~$(e^{i})_{a}$~are left-invariant one-forms, preserved under the
action of~$G$, and the Bianchi time $t$ is proper-time on the
congruence orthogonal to the surfaces of homogeneity.\\

When field equations are imposed on a spatially-homogeneous
space-time, they typically reduce to a system of ordinary
differential equations for the spatial metric $\tilde{a}_{ij}(t)$,
together with some matter equations.

\subsection{Conformal gauge fixing for Bianchi types}
Clearly we may assume that the conformal factor~$Z$~for Bianchi types is a
function only of the cosmic time~$t$. There is a natural choice of
this function, different from that made in Lemma 5.1, given by the 
following lemma.\\

\textbf{Lemma~6.1}~If it is assumed that the conformal factor 
satisfies ~$\nabla_{a}Z\neq 0$~at~$t=0$,~then we may
take~$Z=t^{1/2}$, where~$t$~is the Bianchi time.\\

\textit{Proof}. Suppose~$\tilde{g}_{ab}=Z^{2}g_{ab}$,
with~$g_{ab}$~smooth and~$\tilde{g}_{ab}$~a Bianchi type solution of
the EV system. Let~$V^{2}=-g^{ab}\nabla_{a}Z\nabla_{b}Z$, 
with~$V=1$~at~$t=0$~without loss of generality. Define the positive 
function~$g$~by
\begin{equation}g^{2}=\frac{1}{Z^{2}}\int_{0}^{Z}~\frac{s}{V(s)}~ds\end{equation}
Now let~$\bar{g}_{ab}=g^{-2}g_{ab}~,~\bar{Z}=gZ$. One then
gets~$\bar{V}=1/2$,~$\bar{Z}=t^{1/2}$, and it remains to show that
~$g$~is smooth.\\

One must have~$V^{-1}(s)=1+~h(s)$~for some
smooth~$h$~with~$h(0)=0$. Hence $h(s)=sr(s)$ for some smooth~$r$, and
\begin{displaymath}g^{2}=\frac{1}{2}+\frac{1}{Z^{2}}\int_{0}^{Z}s^{2}r(s)~ds
\end{displaymath}
It follows that~$g$~is
continuous. Now~$r(s)=r(0)+sp(s)$~with~$p$~smooth, and hence
\begin{displaymath}
g^{2}=\frac{1}{2}+\frac{1}{3}r(0)Z+\frac{1}{Z^{2}}\int_{0}^{Z}s^{3}p(s)~ds
\end{displaymath}
It now follows that~$g$~is ~$C^{1}$, and continuing in this way, one
shows that~$g$~is smooth.

\subsection{Evolution equations in M}
We now wish to write the conformal Einstein-Vlasov equations as an
evolution system for the unphysical metric~$g_{ab}$~and the particle
distribution function~$f$~on a manifold~$M=[0,T]\mathbf{\times}G$. We
assume that~$f=f(Z,~v_{i})$~, where the~$v_{i}$~are components of
4-momentum in the invariant frame~$(e^{i})_{a}$.\\
Note that the metric~$g_{ab}$~in~$M$~takes the form
\begin{equation}g_{ab}=-4\nabla_{a}Z\nabla_{b}Z+a_{ij}(Z)(e^{i})_{a}(e^{j})_{b}\end{equation}
with ~$a_{ij}$~positive definite.
\\
The Vlasov equation can be written
\begin{equation}
\frac{\partial f}{\partial Z}=\partial_{Z}f
=2(b^{rs}v_{r}v_{s})^{-1/2}b^{ln}v_{k}v_{l}C^{k}_{jn}
\frac{\partial f}{\partial v_{j}}
\end{equation}
where ~$(b^{ij})=(a_{ij})^{-1}$~and~$C^{i}_{jk}$~are the structure constants of
the Lie group G.
The spatial components of the Einstein equations, in the
basis~$(e^{i})_{a}$~are
\begin{equation}\partial_{Z}(a_{ij})=k_{ij}\end{equation}
\begin{equation}\partial_{Z}(b^{ij})=-b^{in}b^{jm}k_{mn}\end{equation}
\begin{displaymath}\partial_{Z}(k_{ij})=4C^{k}_{ck}(C^{r}_{tj}a_{ir}+a_{jr}C^{r}_{ti})b^{ct}+4C^{c}_{ki}(C^{k}_{cj}+a_{cm}b^{kt}C^{m}_{tj})\end{displaymath}
\begin{displaymath}+2C^{m}_{sk}C^{r}_{tc}a_{jm}a_{ir}b^{kt}b^{cs}+\frac{2}{Z^{2}}\Bigg\{\frac{32\pi}{(\mathrm{det}~a)^{1/2}}\int\frac{fv_{i}v_{j}}{(v^{T}bv)^{1/2}}~d^{3}v~-a_{ij}\Bigg\}\end{displaymath}
\begin{equation}-\frac{2}{Z}k_{ij}-\Big(\frac{1}{Z}a_{ij}+\frac{1}{2}k_{ij}\Big)(b^{rs}k_{rs})+b^{lq}k_{il}k_{qj}
\end{equation}
We wish to remove the second-order pole in $Z$ in (60), so we introduce a new 
independent variable~$Z_{ij}$~according to
\begin{equation}Z_{ij}=\frac{1}{Z}\Bigg\{\frac{32\pi}{(\mathrm{det}~a)^{1/2}}\int\frac{fv_{i}v_{j}}{(v^{T}bv)^{1/2}}~d^{3}v-~a_{ij}\Bigg\}\end{equation}
Then (60) becomes
\begin{displaymath}\partial_{Z}(k_{ij})=4C^{c}_{kc}(C^{r}_{tj}a_{ir}+a_{jr}C^{r}_{ti})b^{kt}+4C^{c}_{ki}(C^{k}_{cj}+a_{cm}b^{kt}C^{m}_{tj})\end{displaymath}
\begin{equation}+2C^{m}_{sk}C^{r}_{tc}a_{jm}a_{ir}b^{kt}b^{cs}+\frac{1}{Z}(2Z_{ij}-2k_{ij}-a_{ij}(b^{rs}k_{rs}))-\frac{1}{2}k_{ij}(b^{rs}k_{rs})+b^{lq}k_{il}k_{qj}\end{equation}
and using (57) we get an evolution equation for~$Z_{ij}$~:
\begin{displaymath}\partial_{Z}(Z_{ij})=\frac{1}{Z}\Bigg\{-Z_{ij}-k_{ij}-\frac{64\pi}{(\mathrm{det}~a)^{1/2}}C^{k}_{np}b^{ln}\int\frac{\partial
f}{\partial
v_{p}}\frac{v_{i}v_{j}v_{k}v_{l}}{(v^{T}bv)}~d^{3}v\end{displaymath}
\begin{equation}+\frac{16\pi}{(\mathrm{det}~a)^{1/2}}(b^{en}b^{fm}k_{mn}-b^{ef}b^{rs}k_{rs})\int
\frac{f~v_{i}v_{j}v_{e}v_{f}}{(v^{T}bv)^{3/2}}~d^{3}v~\Bigg\}\end{equation}
Note that if (57)-(59), (62)-(63) are solved, and we define
~$\bar{Z}_{ij}$~by the rhs of (61) then one gets
\begin{displaymath}\partial_{Z}(\bar{Z}_{ij}-Z_{ij})=-\frac{1}{Z}(\bar{Z}_{ij}-Z_{ij})\end{displaymath}
and hence~$\bar{Z}_{ij}=Z_{ij}$~, and the definition (61) is
recovered. 
Also, from (58)-(59) there follows
\begin{displaymath}
\partial_{Z}(a_{ij}b^{jk}-\delta_{i}^{k})=
-(a_{ij}b^{js}-\delta_{i}^{s})k_{sn}b^{kn}
\end{displaymath}
so that, by a Gronwall estimate,
if~$a_{ij}b^{jk}=\delta_{i}^{k}$~at~$Z=0$~,
then~$a_{ij}b^{jk}=\delta_{i}^{k}$~for all $Z$. 
\subsection{Constraints and initial data}
The Hamiltonian constraint is
\begin{displaymath}\frac{16\pi}{(\mathrm{det}~a)^{1/2}}\int
f(b^{rs}v_{r}v_{s})^{1/2}~d^{3}v=\frac{3}{2}+~^{(3)}R~Z^{2}-\left(\frac{Z^{2}}{16}\right)b^{ip}b^{jq}k_{pq}k_{ij}\end{displaymath}
\begin{equation}+\left(\frac{Z}{2}\right)b^{ip}b^{jq}k_{pq}a_{ij}+\frac{Z^{2}}{16}(b^{rs}k_{rs})^{2}\end{equation}
where~$^{(3)}R$~is the Ricci scalar of~$a_{ij}$.
\\The momentum constraint is
\begin{equation}\frac{32\pi}{(\mathrm{det}~a)^{1/2}}\int
fv_{i}~d^{3}v=Z^{2}(b^{pl}k_{lj}C^{j}_{pi}+b^{sl}k_{li}C^{r}_{sr})\end{equation}
Suppose that (57)-(59), (62)-(63) are satisfied, and write
(64)-(65) as~$C=0,~C_{i}=0$~respectively. Then one calculates that
\begin{displaymath}\partial_{Z}(Z^{2}(\mathrm{det}~a)C)=0~~,~~\partial_{Z}C_{i}=-\frac{1}{2}(b^{rs}k_{rs})C_{i}\end{displaymath}
Hence~$C\equiv0$~and, by a Gronwall
estimate,~$C_{i}\equiv0$~if~$C_{i}=0$~initially.
\\
To satisfy the momentum constraint at~$Z=0$~one must have
\begin{equation}
\int~f^{0}v_{i}~d^{3}v=0
\end{equation}
which is the counterpart of (22), and if the evolution equations 
are to be satisfied, then from (61)
\begin{equation}
a^{0}_{ij}=\frac{32\pi}{(\mathrm{det}~a^{0})^{1/2}}\int\frac{f^{0}v_{i}v_{j}}{(v^{T}b^{0}v)^{1/2}}~d^{3}v
\end{equation}
which is the counterpart of (23) (recall here $V^0=1/2$). If (67) is satisfied, then by taking
the trace we see that the
Hamiltonian constraint is also satisfied initially.\\

To determine~$k^{0}_{ij},~Z^{0}_{ij}$, note that from (62)-(63)
\begin{equation}Z^{0}_{ij}-k^{0}_{ij}-\frac{1}{2}a^{0}_{ij}((b^{0})^{rs}k^{0}_{rs})=0\end{equation}
and
\begin{displaymath}Z^{0}_{ij}+k^{0}_{ij}+\frac{64\pi}{(\mathrm{det}~a^{0})^{1/2}}C^{k}_{np}(b^{0})^{ln}\int\frac{\partial
f^{0}}{\partial v_{p}}\frac{v_{i}v_{j}v_{k}v_{l}}{(v^{T}b^{0}v)}~d^{3}v\end{displaymath}
\begin{equation}+\frac{16\pi}{(\mathrm{det}~a^{0})^{1/2}}(-(b^{0})^{en}(b^{0})^{fm}k^{0}_{mn}+(b^{0})^{ef}(b^{0})^{rs}k^{0}_{rs})\int\frac{f^{0}v_{i}v_{j}v_{e}v_{f}}{(v^{T}b^{0}v)^{3/2}}~d^{3}v~=0\end{equation}

Now eliminate~$Z^{0}_{ij}$~to get
\begin{equation}2k^{0}_{ij}+~\mathrm{tr}k^{0}(a^{0}_{ij})-(k^{0})^{ef}\chi_{ijef}+\frac{64\pi}{(\mathrm{det}~a^{0})^{1/2}}C^{k}_{np}(b^{0})^{ln}\int\frac{\partial
f^{0}}{\partial
v_{p}}\frac{v_{i}v_{j}v_{k}v_{l}}{(v^{T}b^{0}v)}~d^{3}v=0\end{equation}
where
\begin{equation}\chi_{ijkl}=\frac{16\pi}{(\mathrm{det}~a^{0})^{1/2}}\int\frac{f^{0}v_{i}v_{j}v_{k}v_{l}}{(v^{T}b^{0}v)^{3/2}}~d^{3}v~~~,~~~\mathrm{tr}k^{0}=(b^{0})^{rs}k^{0}_{rs}\end{equation}
Taking the trace of (70) gives
\begin{equation}0=~\frac{9}{2}\mathrm{tr}k^{0}+~\frac{64\pi}{(\mathrm{det}~a)^{1/2}}C^{k}_{np}(b^{0})^{ln}\int~\frac{\partial
f^{0}}{\partial v_{p}}v_{k}v_{l}~d^{3}v\end{equation}
and then (66) implies~that~$\mathrm{tr}k^{0}=0$.
Equation (70) now becomes
\begin{equation}
(\chi_{ijef}-2a^{0}_{ie}a^{0}_{jf})(k^{0})^{ef}=-\frac{128\pi}{(\mathrm{det}~a^{0})^{1/2}}(b^{0})^{ln}C^{k}_{n(i}\int\frac{f^{0}}{(v^{T}b^{0}v)}v_{j)}v_{k}v_{l}~d^{3}v
\end{equation}
which corresponds to (43). By the eigenvalue properties
of~$\chi_{ijkl}$ noted in section 4.3,~$k^{0}_{ij}$~is
uniquely determined by~$f^{0}$,~and $Z^{0}_{ij}=k^{0}_{ij}$~by (68).\\

In summary, the free data at~$Z=0$~consists of
just~$f^{0}(v_{i})$~satisfying the integral constraint (66),
with$~a^{0}_{ij},~k^{0}_{ij}=Z^{0}_{ij}$~being determined by (67),
(73) respectively. Given such data the constraints are preserved by
the evolution, and hence the task is just to solve (57)-(59),
(62)-(63).
\subsection{Solving the evolution equations}
We will now use Theorem 1 of (Rendall and Schmidt 1991), together with an
 iteration technique due to (Rendall 1994), to solve the
 conformal Einstein-Vlasov equations near~$Z=0$. The result, after a
 series of lemmas, is Theorem 6.1 below.\\

Choose~$f^{0}\in~C_{0}^{\infty}(\mathbb{R}^{3})$~,~$f^{0}\geq 0$,~with ~$f^{0}$~supported
outside some open ball centred on the origin, and
let~$a^{0}_{ij},~k^{0}_{ij}=Z^{0}_{ij}$~be determined as above.
\\Define
\begin{displaymath}m_{ij}=k_{ij}-k^{0}_{ij}~~~,~~~w_{ij}=Z_{ij}-Z^{0}_{ij}\end{displaymath}
so that~$m^{0}_{ij}=w^{0}_{ij}=0$.
\\The evolution equations can now be written as follows
\begin{equation}
\partial_Zf=2(b^{rs}v_{r}v_{s})^{-1/2}b^{ln}v_{k}v_{l}C^{k}_{jn}\frac{\partial
f}{\partial v_{j}}\end{equation}
\begin{equation}\partial_{Z}(a_{ij})=k^{0}_{ij}+~m_{ij}\end{equation}
\begin{equation}\partial_{Z}(b^{ij})=-b^{in}b^{jm}(k^{0}_{mn}+m_{mn})\end{equation}
\begin{displaymath}\partial_{Z}(m_{ij})=4C^{c}_{kc}(C^{r}_{tj}a_{ir}+a_{jr}C^{r}_{ti})b^{kt}+4C^{c}_{ki}(C^{k}_{cj}+a_{cm}b^{kt}C^{m}_{tj})\end{displaymath}
\begin{displaymath}+2C^{m}_{sk}C^{r}_{tc}a_{jm}a_{ir}b^{kt}b^{cs}-\frac{1}{2}(k^{0}_{ij}+m_{ij})b^{mn}(k^{0}_{mn}+m_{mn})+b^{lq}(k^{0}_{il}+m_{il})(k^{0}_{qj}+m_{qj})\end{displaymath}
\begin{equation}+P_{ij}^{~~mn}(k^{0}_{mn}+m_{mn})+~\frac{1}{Z}\{-a^{0}_{ij}(b^{0})^{mn}m_{mn}+2w_{ij}-2m_{ij}\}\end{equation}
\begin{displaymath}\partial_{Z}(w_{ij})=~S^{(1)}_{ij}+~S^{(2)}_{ij}+~\frac{1}{Z}\Bigg\{-w_{ij}-m_{ij}\end{displaymath}
\begin{equation}+\Bigg(\frac{16\pi}{(\mathrm{det}~a^{0})^{1/2}}\int\frac{f^{0}v_{i}v_{j}v_{e}v_{f}}{(v^{T}b^{0}v)^{3/2}}~d^{3}v\Bigg)((b^{0})^{en}(b^{0})^{fm}m_{mn}-(b^{0})^{ef}(b^{0})^{mn}m_{mn})\Bigg\}\end{equation}
where
\begin{equation}P_{ij}^{~~mn}(t)=\int_{0}^{1}\{a_{ij}b^{pm}b^{qm}(k^{0}_{pq}+m_{pq})-b^{mn}(k^{0}_{ij}+m_{ij})\}(st)~ds\end{equation}
and
\begin{displaymath}S^{(1)}_{ij}(t)=\int_{0}^{1}-\frac{64\pi}{(a)^{1/2}}C^{k}_{np}\Bigg\{(b^{rs}(k^{0}_{rs}+m_{rs})b^{ln}-b^{lp}b^{nq}(k^{0}_{pq}+m_{pq}))\int\frac{\partial
f}{\partial v_{p}}\frac{v_{i}v_{j}v_{k}v_{l}}{(v^{T}bv)}~d^{3}v\end{displaymath}
\begin{displaymath}+b^{ln}b^{ru}b^{sv}(k^{0}_{uv}+m_{uv})\int\frac{\partial
f}{\partial
v_{p}}\frac{v_{i}v_{j}v_{k}v_{l}v_{r}v_{s}}{(v^{T}bv)^{2}}~d^{3}v\end{displaymath}
\begin{equation}+2C^{w}_{qr}b^{rs}b^{ln}\int\frac{\partial}{\partial
v_{p}}\Bigg(\frac{v_{w}v_{s}}{(v^{T}bv)^{1/2}}\frac{\partial
f}{\partial
v_{q}}\Bigg)\frac{v_{i}v_{j}v_{k}v_{l}}{(v^{T}bv)}~d^{3}v\Bigg\}(st)~ds\end{equation}
\begin{displaymath}S^{(2)}_{ij}(t)=(k^{0}_{mn}+m_{mn})\int_{0}^{1}\frac{16\pi}{(a)^{1/2}}\Bigg\{2C^{k}_{pr}b^{lr}\int\frac{\partial
f}{\partial
v_{p}}\frac{v_{i}v_{j}v_{e}v_{f}v_{k}v_{l}}{(v^{T}bv)^{2}}~d^{3}v\end{displaymath}
\begin{displaymath}+(k^{0}_{rs}+m_{rs})\Bigg[\frac{3}{2}b^{pr}b^{qs}(b^{en}b^{fm}-b^{ef}b^{mn})\int\frac{fv_{i}v_{j}v_{e}v_{f}v_{p}v_{q}}{(v^{T}bv)^{5/2}}~d^{3}v\end{displaymath}
\begin{displaymath}+\Bigg(\int\frac{fv_{i}v_{j}v_{e}v_{f}}{(v^{T}bv)^{3/2}}~d^{3}v\Bigg)\Big(b^{rs}(b^{en}b^{fn}-b^{ef}b^{mn})\end{displaymath}
\begin{equation}-b^{er}b^{ns}b^{fm}-b^{en}b^{fr}b^{ms}+b^{mn}b^{er}b^{fs}+b^{ef}b^{mr}b^{ns}\Big)\Bigg]\Bigg\}(st)~ds\end{equation}
where~$a=$det$(a_{ij})$.\\

Suppose now that ~$f^{0}$~is supported outside a ball of radius
~$B_{1}$~centred on the origin. Let~$r=(v^{T}v)^{1/2}$,~and
let~$\phi(v_{i})$~be a smooth function such
that~$\phi=1$~on~$r>B_{2}$,~$B_{2}<B_{1}$,~$\phi=0$~on~$r<B_{3}$~,
~$B_{3}<B_{2}$,~and~$0\leq\phi\leq 1$~elsewhere. Replace the Vlasov
equation (57) by
\begin{equation}\frac{\partial
f}{\partial Z}=2\phi(b^{rs}v_{r}v_{s})^{-(1/2)}b^{ln}v_{k}v_{l}C^{k}_{jn}\frac{\partial
f}{\partial v_{j}}\end{equation}
Equations (77)-(78) can be written
\begin{equation}\partial_{Z}(u)+\frac{1}{Z}Nu=G(a,b,u,f)\end{equation}
where~$u$~stands for ~$m,~w$,~and ~$N$~is a constant matrix containing
just~$a^{0}_{ij},~(b^{0})^{ij}$.\linebreak $N$~has desirable properties
with regard to application of the theorem of Rendall and Schmidt
(1991), given by
the following lemma:\\

\textbf{Lemma~6.2}~All the eigenvalues of~$N$~have positive real part
and~$N$~is diagonalisable.\\

\textit{Proof}.~The eigenvalue equation for~$N$~is 
\begin{equation}2m_{ij}+ma^{0}_{ij}-2w_{ij}=\lambda
m_{ij}\end{equation}
\begin{equation}m_{ij}+\frac{1}{2}ma^{0}_{ij}+w_{ij}-\chi_{ijkl}m^{kl}=\lambda
w_{ij}\end{equation}
where~$m_{ij},~w_{ij}$~are components of an eigenvector, 
~$m=(b^{0})^{ij}m_{ij}$, and $\lambda$ is the eigen-value.\\

Hence~$w_{ij}=\frac{1}{2}(2-\lambda)m_{ij}+\frac{1}{2}ma^{0}_{ij}$,~and
substituting in (85) gives
\begin{equation}(\lambda^{2}-3\lambda+4)m_{ij}-(\lambda-2)ma^{0}_{ij}=2\chi_{ijkl}m^{kl}\end{equation}
We know from (67) that~$(b^{0})^{ij}\chi_{ijkl}=\frac{1}{2}a^{0}_{kl}$,~and so
taking the trace of (86) gives
\begin{equation}(\lambda-3)^{2}m=0\end{equation}
If~$m\neq 0$~then$~\lambda=3$~and an eigenvector
is~$w_{ij}=m_{ij}=\frac{1}{3}ma^{0}_{ij}$~. If $\lambda\neq
3$ then~$m=0$~and (86) becomes
\begin{equation}(\lambda^{2}-3\lambda+4)m^{ij}=2m^{kl}\chi^{ij}_{~~kl}\end{equation}
and we must consider eigenvalues of$~\chi^{ij}_{~~kl}$,~which we may regard
as a~$9\times9$~symmetric matrix, with 9 real eigenvalues and 9
linearly independent eigenvectors. So suppose
\begin{equation}\chi^{ij}_{~~kl}m^{kl}=\mu m^{ij}\end{equation}
Then~$m_{ij}=m_{ji}$~and we may work in an invariant basis where
~$a^{0}_{ij}=\delta_{ij},~(m_{ij})=\mathrm{diag}(m_{1},~m_{2},~m_{3})$.
Clearly~$\mu$~must be positive since
\begin{displaymath}\chi_{ijkl}m^{ij}m^{kl}\geq 0\end{displaymath}
It follows that
\begin{displaymath}\mu|m_{i}|=16\pi\Big|\int\frac{f^{0}}{(v^{T}v)^{3/2}}v_{i}^{2}(v_{1}^{2}m_{1}+v_{2}^{2}m_{2}+v_{3}^{2}m_{3})~d^{3}v\Big|\end{displaymath}
\begin{displaymath}\leq16\pi\int\frac{f^{0}v_{i}^{2}}{(v^{T}v)^{3/2}}(v_{1}^{2}|m_{1}|+v_{2}^{2}|m_{2}|+v_{3}^{2}|m_{3}|)~d^{3}v\end{displaymath}
and hence
\begin{displaymath}\mu\sum_{i=1}^{3}|m_{i}|\leq16\pi\int\frac{f^{0}}{(v^{T}v)^{1/2}}(v_{1}^{2}|m_{1}|+v_{2}^{2}|m_{2}|+v_{3}^{2}|m_{3}|)~d^{3}v~=\frac{1}{2}\sum_{i=1}^{3}|m_{i}|\end{displaymath}
by (67), so that~$\mu\leq \frac{1}{2}$.
\\Going back to (88) one gets
that~$\lambda=\frac{1}{2}(3\pm[8\mu-7]^{1/2})$~with~$0\leq\mu\leq\frac{1}{2}$.~Thus~$\lambda$~
has positive real part, and there exist two~$\lambda 's$~for each of
the nine linearly independent~$m_{ij}$~. One has
~$w_{ij}=\frac{1}{2}(2-\lambda)m_{ij}$, and this gives 18 linearly
independent eigenvectors of~$N$.
\\\\
Lemma 6.2 gives that there exists an 18$\times$18 matrix~$L$~ such that
equations (83) may be written
\begin{equation}\partial_{Z}(y_{\alpha})+\frac{1}{Z}\lambda_{\alpha}y_{\alpha}=~H_{\alpha}(a,b,y,f)\end{equation}
where
~$y=Lu,~H(a,b,y,f)=LG(a,b,L^{-1}y,f)$~,
and~$\lambda_{\alpha}$~are the eigenvalues of~$N$.
\\
The equations to be solved are now (75), (76), (82), (90).\\

Consider first the modified Vlasov equation (82). The
characteristics of this equation are defined as the
solutions~$V_{j}(s,t,v)$~of the system
\begin{equation}\frac{dV_{j}}{ds}=-2\phi(V)(b^{rs}V_{r}V_{s})^{-1/2}b^{ln}V_{k}V_{l}C^{k}_{jn}\end{equation}
with~$V_{j}(t,t,v)=v_{j}$. Then the solution of (82) is given
by~$f(t,v_{i})=f^{0}(V_{j}(0,t,v_{i}))$~, and is smooth by standard theory.\\
\\Now define an iteration as follows: Let
~$f^{0},~a^{0}_{ij},~(b^{0})^{ij}$~be the initial data as above, and
let~$m^{0}_{ij}=w^{0}_{ij}=0$. If smooth iterates~$a^{n},~b^{n},~m^{n},~w^{n},~f^{n}$~are given for some~$n$,~let~$y^{n}=Lu^{n}$~, and determine~$V^{n+1}$~by solving (91) with~$b^{n}$~in the rhs, and let~$f^{n+1}(t,v)=f^{0}(V^{n+1}(0,t,v))$~which solves 
the $n$-th Vlasov equation. Now
substitute~$a^{n},~b^{n},~y^{n},~f^{n+1}$~ into the rhs of (75),
(76), (90). By Theorem 1 of (Rendall and Schmidt 1991) these linear ODE's have
a unique smooth solution~$a^{n+1},~b^{n+1},~y^{n+1}$~on some small
time interval~$[0,T'_{n+1})$. Let~$[0,T_{n+1})$~be the largest time
interval on which~$(\mathrm{det}~a^{n})$~remains strictly positive. Then
by standard theory for regular, linear ODE's, the (n+1)th solution can
be extended uniquely and smoothly onto~$[0,T_{n+1})$. 
\\\\
Let~$|a|$~be the maximum modulus of any component of~$(a_{ij})$~, with
a similar definition for other quantities, and suppose that~$\forall
n<N$~the following bounds hold:
\begin{displaymath}|a^{n}-a^{0}|\leq A_{1}~~~~~~|b^{n}-b^{0}|\leq
A_{2}\end{displaymath}
\begin{displaymath}|m^{n}|\leq A_{3}~~~~~~|w^{n}|\leq
A_{4}\end{displaymath}
\begin{equation}(\mathrm{det}~b^{n})^{-1}< A_{6}~~(A_{6}>~(\mathrm{det}~b^{0})^{-1})\end{equation}
Suppose also that ~$f^{n}(t,v)\neq 0~\Rightarrow~(v^{T}v)^{1/2}\leq
A_{5}$
\\Now the speed of propagation for the nth Vlasov equation, at the
point~$v_{i}$~will be bounded by~$C_{1}(v^{T}v)^{1/2}$~, with
~$C_{1}=C_{1}(A_{2},A_{6})$. Hence if\linebreak~$R=\mathrm{diam~supp}~f^{N}(t)$~,
then
\begin{equation}\frac{dR}{dt}\leq
C_{1}R~~\Rightarrow~~R\leq~P_{0}\mathrm{exp}(C_{1}t)\end{equation}
where $~P_{0}=~\mathrm{diam~supp}~f^{0}$.
\\Now (75),~(76) give
\begin{equation}|a^{N}-a^{0}|\leq C_{2}t~~~~~~|b^{N}-b^{0}|\leq
C_{3}t\end{equation}
and then one gets
\begin{equation}(\mathrm{det}~b^{N})^{-1}<A_{6}\end{equation}
for~$t$~in a small enough
interval~$[0,T]$, where~$T=T(A_{i})$. Since~$f^{0}$~is bounded one also
gets:~$|H^{n}|\leq C_{4}(A_{i})$~for~$n<N$.\\

From (90) there follows
\begin{equation}y^{N}_{\alpha}(t)=t^{-\lambda_{\alpha}}\int_{0}^{t}s^{\lambda_{\alpha}}H^{N-1}_{\alpha}(s)~ds\end{equation}
and hence~$|y^{N}_{\alpha}(t)|\leq C_{4}t$~. Also, from the bound on the
speed of propagation, if~$f^{0}$~is supported outside~$r=B_{1}$~,
then~$f^{N}$~is supported outside\linebreak~$r=B_{1}\mathrm{exp}(-C_{1}t)$.
\\
Now choose~$A_{5}>P_{0}$~, and choose~$T=T(A_{i},C_{j})$~small enough
so that ~$R<A_{5}$, and for ~$t\in [0,T]~~f^{N}(t)$~is supported
on~$\{v:~\phi(v)=1\}$, and the following bounds hold:
\begin{displaymath}|a^{N}-a^{0}|\leq A_{1}~~~~~~|b^{N}-b^{0}|\leq
A_{2}\end{displaymath}
\begin{equation}|m^{N}|\leq A_{3}~~~~~~|w^{N}|\leq A_{4}\end{equation}
Then by induction all iterates exist, are smooth, and are uniformly
bounded on~$[0,T]$. Also,~$\forall n,~f^{n}$~is supported
on~$\{v:\phi(v)=1\}$, and~$\mathrm{diam~supp}(f^{n})\leq A_{5}$.
\\
For the difference between successive iterates one has the following
estimates:
\begin{equation}|a^{n+1}-a^{n}|(t)\leq
C\int_{0}^{t}~|y^{n}-y^{n-1}|(s)~ds\end{equation}
\begin{equation}|b^{n+1}-b^{n}|(t)\leq
C\int_{0}^{t}~|b^{n}-b^{n-1}|(s)+|y^{n}-y^{n-1}|(s)~ds\end{equation}
\begin{displaymath}|y^{n+1}-y^{n}|(t)\leq
C\int_{0}^{t}\Big\{(|a^{n}-a^{n-1}|+|b^{n}-b^{n-1}|+|y^{n}-y^{n-1}|)(s)\end{displaymath}
\begin{equation}+\Big(\int_{0}^{1}~(|a^{n}-a^{n-1}|+|b^{n}-b^{n-1}|+|y^{n}-y^{n-1}|+\|f^{n+1}-f^{n}\|_{\infty})(sp)~dp\Big)\Big\}~ds\end{equation}
Equation (98) comes from (75), (99) comes from (76), and
(100) comes from (96).\\\\

From the characteristic equation (91) one gets an inequality similar
to that in (93) and hence the~$V^{n}$~are uniformly bounded, above
and below for
~$v$~in a compact set with~$\phi(v)=1$~, and~$t\leq T,~0\leq s \leq t$. Now suppose~$v\in
\{v:\phi(v)=1\}\cap\{v:(v^{T}v)^{1/2}<A_{5}\}$. Then from (91) there follows the estimate:
\begin{equation}\Big|\frac{dV^{n+1}}{ds}-\frac{dV^{n}}{ds}\Big|(s,t,v)\leq
C\{|V^{n+1}-V^{n}|+|b^{n}-b^{n-1}|\}\end{equation}
for~$t\leq T,~0\leq s\leq t$.
\\
Define the quantity~$\alpha^{n}(t)$~by
\begin{displaymath}\alpha^{n}(t)=~\mathrm{sup}\{|V^{n+1}-V^{n}|(s,t,v):~0\leq
s\leq
t,~r<A_{5},~\phi(v)=1\}\end{displaymath}
\begin{equation}+|a^{n+1}-a^{n}|(t)+|b^{n+1}-b^{n}|(t)+|y^{n+1}-y^{n}|(t)\end{equation}
Then
\begin{equation}\|f^{n+1}-f^{n}\|_{\infty}(t)\leq~\|f^{0}\|_{C^{1}}(\alpha^{n}(t))\end{equation}
Combining the above inequalities gives
\begin{equation}\alpha^{n}(t)\leq~C\int_{0}^{t}\Big\{\alpha^{n}(s)+\alpha^{n-1}(s)+\int_{0}^{1}(\alpha(sp)+\alpha^{n-1}(sp))~dp\Big\}~ds\end{equation}
Now Gronwall's inequality implies
\begin{equation}\alpha^{n}(t)\leq~C\int_{0}^{t}\Big\{\alpha^{n-1}(s)+\Big(\int_{0}^{1}\alpha^{n}(sp)+\alpha^{n-1}(sp)~dp\Big)\Big\}~ds\end{equation}
for~$t\in~[0,T]$, and since the rhs is increasing , this implies
\begin{displaymath}\mathrm{sup}_{(0\leq s\leq t)}\alpha^{n}(s)~\leq
C\int_{0}^{t}\Big\{\alpha^{n-1}(s)+\Big(\int_{0}^{1}(\alpha^{n}(sp)+\alpha^{n-1}(sp))~dp\Big)\Big\}ds\end{displaymath}
\begin{equation}\leq~C_{1}t\Big(\mathrm{sup}_{(0\leq s\leq
t)}\alpha^{n}(s)\Big)+~C_{2}\int_{0}^{t}\mathrm{sup}_{(0\leq p\leq
s)}(\alpha^{n-1}(p))~ds\end{equation}
Put~$\beta^{n}(t)=~\mathrm{sup}_{(0\leq s\leq t)}\alpha^{n}(s)$. Then
(106) is just
\begin{equation}\beta^{n}(t)(1-C_{1}t)\leq~C_{2}\int_{0}^{t}~\beta^{n-1}(s)~ds\end{equation}
Choose~$T_{1}\leq T$~so that~$(1-C_{1}t)\geq \frac{1}{2}$~for~$t\in
[0,T_{1}]$. It follows now that
\begin{equation}\beta^{n}(t)\leq
2C_{2}\int_{0}^{t}~\beta^{n-1}(s)~ds\end{equation}
and hence
\begin{equation}\beta^{n}(t)\leq~C^{n-2}\|\beta^{2}\|_{\infty}t^{n-2}/(n-2)!\end{equation}
Therefore~~$V^{n},~b^{n},~a^{n},~y^{n}$~are uniformly Cauchy
on~$[0,T_{1}]$. It follows that~$a^{n},~b^{n},~y^{n}$~have 
continuous limits~$a(t),~b(t),~y(t)$
~on~$[0,T_{1}]$, and\linebreak[4]$V^{n}(s,t,v)$~has a limit~$V(s,t,v)$~continuous
in~$s,t$. From (75), (76), one gets that~$a,b$~are~$C^{1}$, and
from (91)~$V(s,t,v)$~is~$C^{1}$~in ~$s,t$, so that~$a,b,V$~in fact
satisfy these equations. It is then standard (Taylor 1996) that ~$V$~is
~$C^{1}$~in~$v$. If we put~$f(t,v)=f^{0}(V(0,t,v))$,
then~$f$~is supported on~$\{v:\phi(v)=1\}$~and hence satisfies the
Vlasov equation (74) .\\
Note that we can differentiate (91), and so~$V$~is~$C^{2}$~along
with~$f$.\\

Clearly from (90), (75), (76), (91) we get uniform convergence
of the time derivatives of all iterates on~$[\epsilon,T_{1}]$~for
arbitrary~$0<\epsilon<T_{1}$~, and inductively it follows that all the
sequential limits are smooth on~$(0,T_{1}]$.
\\\\
If we let~$u=(m,w)=L^{-1}y$~as before, then~$u$~is continuous
on~$[0,T_{1}]$~, smooth
on~$(0,T_{1}]$~, and satisfies (83)
there.
Putting~$k_{ij}=k^{0}_{ij}+m_{ij},~Z_{ij}=Z^{0}_{ij}+w_{ij}$~gives
that~$k_{ij},~Z_{ij}$~are continuous on~$[0,T_{1}]$~, smooth
on~$(0,T_{1}]$~, and satisfy (62), (63) there. These equations can
be written
\begin{equation}\frac{dx}{dZ}+\frac{1}{Z}N(a,b,f)x=G_{1}(a,b,x)+\frac{1}{Z}G_{2}(a,b,f)\end{equation}
where~$x$~stands for~$k_{ij},~Z_{ij}$~,~$G_{1}$~is quadratic in~$x$,
 ~$G_{2}$~accounts for the terms~$64\pi C^{k}_{np}$\ldots ,
 and~$Z\in (0,T_{1}]$.\\

Now regard~$a,b,f$~as known~$C^{1}$~functions on~$[0,T_{1}]$. Then
 (110) implies
\begin{equation}Z\frac{dx}{dZ}+Nx=(N-N(Z))x+Z
 F_{1}(Z,x)+F_{2}(Z)\end{equation}
where ~$N$~is as in (83),~$F_{1}$~is~$C^{1}$~in~$Z$~and
quadratic in~$x$, and~$F_{2}$~is~$C^{1}$. We can now differentiate
(111) on~$(0,T_{1}]$~to get
\begin{equation}Z\frac{dq}{dZ}+(I+N)q=Z R(Z)q+S(Z)\end{equation}
where~$q=\frac{dx}{dZ}$~and~$R,~S$~are continuous functions
on~$[0,T_{1}]$.\\

It follows by a lemma of (Rendall and Schmidt 1991) that (112) has a
unique~$C^{1}$~solution on~$(0,T_{1}]$, and that this extends to a
continuous solution of the associated integral equation
on~$[0,T_{1}]$. Hence~$x$~is~$C^{1}$, and~$(a,~b,~k,~z,~f)$\linebreak[4]is a
classical solution of the conformal Einstein-Vlasov equations. Inductively one gets that this solution is in fact~$C^{\infty}$. \\\\If
two solutions with the same initial data are given, then it is
possible to derive an inequality for their difference similar to that
obtained for the quantity~$\beta^{n}$. It follows that the solution
constructed above is unique.
\\\\
We have therefore proved the following:
\\\\
\textbf{Theorem 6.1.}~Let~$G$~be a 3-dimensional Lie group of some Bianchi
type~$n$, and let~$(e^{i})_{a}$~be a basis of left-invariant one-form fields on~$G$~. Suppose we are given a smooth function~$F^{0}$~on the cotangent
bundle of~$G$~such that~$F^{0}(x,p)=f^{0}(p_{i})$~where~$p_{i}$~are the
components of~$p$~in the frame~$(e^{i})_{a}$. Suppose also
that~$f^{0}$~is compactly supported, supported outside a
neighbourhood of the origin, and the integral constraint
(66) holds. Then there exists exactly one smooth Bianchi
type~$n$~ solution~$(\tilde{f}(t,p_{i}),\tilde{g}_{ij}(t))$~of the
massless Einstein-Vlasov equations on~$G\times
(0,T]$~with an isotropic singularity,
satisfying~$\tilde{f}(t,p_{i})\rightarrow
f^{0}(p_{i})$~as~$t\rightarrow 0$~. 
\\\\
Note that since the initial 3-metric~$a^{0}_{ij}$~ is determined
by~$f^{0}$~only through the integral relation (67) there will be
many (anisotropic)~$f^{0}$ which give rise to the same
~$a^{0}_{ij}$. But then ~$k^{0}_{ij}$~is determined by the third and
fourth moments of~$f^{0}$, and so we expect in general to be able to get
different values of ~$k^{0}_{ij}$~, and thus different 4-geometries,
from the same starting metric~$a^{0}_{ij}$. This is in contrast to the
perfect fluid cosmologies of ATI, where it was shown that the 4-geometry is in 1-1 correspondence with the 3-geometry on~$\Sigma$.
\subsection{FRW data}
The Bianchi types I, V and IX contain, respectively, the~$k=0, -1$~and~$
+1$\linebreak[4] FRW models. We now discuss the question of which~$f^{0}$~constitute FRW data for the conformal EV equations in these
Bianchi types.
\\\\
\textbf{(a) Type I}\\
The situation here is very simple, as all the
~$C^{i}_{jk}$~are zero, and one solution of the field equations
is~$\tilde{f}(Z,p_{i})=f^{0}(p_{i}),~\tilde{g}_{ij}=Z^{2}a^{0}_{ij}$~, for
any admissible~$f^{0}$.
Hence Theorem 6.1 gives that all Bianchi I solutions of massless EV with an isotropic singularity are
in fact FRW and the Weyl tensor is always zero.
\\\\
\textbf{(b) Type V}\\
Here the structure constants are
\begin{equation}C^{i}_{jk}=\delta_{j}^{~i}a_{k}-\delta_{k}^{~i}a_{j}\end{equation}
for some vector~$a_{i}$.
\\All type V 3-metrics have constant (negative) curvature, and for
~$f^{0}$~to be FRW data one must certainly have
also~$k^{0}_{ij}=0$,~since~$\textrm{tr}k^{0}=0$. An isotropic~$f^{0}$~satisfies
this condition by (73), and gives the FRW evolution
~$\tilde{f}=f^{0},~\tilde{g}_{ij}=R^{2}(Z)a^{0}_{ij}$, for some
function~$R$. It is not clear whether~$f^{0}$~\textit{must} be
isotropic to give an FRW solution (see (117) below).\\
Leaving this question aside for the moment, we consider a related 
question, namely whether one can pick an anisotropic~$f^{0}$~so
that the Weyl tensor vanishes at~$\Sigma$~, but the evolution given by
Theorem 6.1 is not FRW. We calculate that the electric and magnetic 
parts of the Weyl tensor defined relative to~$\nabla^{a}Z$~vanish 
at~$Z=0$~if and only if
\begin{equation}D_{m}k^{0}_{l(j}e_{i)}^{~lm}=0\end{equation}
\begin{equation}k^{0}_{ie}(k^{0})^{e}_{~j}+P_{ij}+\lambda a^{0}_{ij}=0\end{equation}
where~$e_{ijk}$ is the volume form, $\lambda$~just takes care of the trace and
\begin{displaymath}P_{ij}\equiv \frac{32\pi}{(a^{0})^{1/2}} C^{s}_{tr}\int \frac{\partial f^{0}}{\partial
v_{t}}\frac{(v^{T}k^{0}v)(b^{0})^{qr}v_{q}v_{s}v_{i}v_{j}}{(v^{T}b^{0}v)^{5/2}}~d^{3}v+\frac{24\pi}{(a^{0})^{1/2}}\int
\frac{f^{0}(v^{T}k^{0}v)^{2}v_{i}v_{j}}{(v^{T}b^{0}v)^{5/2}}~d^{3}v\end{displaymath}
\begin{displaymath}-2(k^{0})^{en}(k^{0})^{f}_{~n}\chi_{ijef}\end{displaymath}
\begin{displaymath}-\frac{64\pi}{(a^{0})^{1/2}}
C^{k}_{np}\Bigg[-(k^{0})^{ln}\int\frac{f^{0}}{(v^{T}b^{0}v)}\left(\{\delta^{p}_{~i}v_{j}v_{k}v_{l}+\textrm{perm}~ijkl\}-\frac{2v_{i}v_{j}v_{k}v_{l}v_{m}(b^{0})^{mp}}{(v^{T}b^{0}v)}\right)d^{3}v
\end{displaymath}
\begin{displaymath}+2C^{s}_{tr}(b^{0})^{ln}\int\frac{\partial f^{0}}{\partial v_{t}}\frac{v_{s}v_{q}(b^{0})^{qr}}{(v^{T}b^{0}v)^{3/2}}\{\delta^{p}_{~i}v_{j}v_{k}v_{l}+\textrm{perm}~ijkl\}~d^{3}v\end{displaymath}
\begin{displaymath}+(b^{0})^{ln}\int\frac{f^{0}}{(v^{T}b^{0}v)^{2}}(v^{T}k^{0}v)\{\delta^{p}_{~i}v_{j}v_{k}v_{l}+\textrm{perm}~ijkl\}~d^{3}v\end{displaymath}
\begin{equation}-2(b^{0})^{ln}\int\frac{f^{0}}{(v^{T}b^{0}v)^{2}}v_{i}v_{j}v_{k}v_{l}v_{m}\left(\frac{2(b^{0})^{mp}(v^{T}k^{0}v)}{(v^{T}b^{0}v)}-(k^{0})^{mp}\right)~d^{3}v\Bigg]
\end{equation}
where~$a^{0}=$det$a^{0}_{ij}$.

Now suppose that one has an FRW Einstein-Vlasov
metric:~$a_{ij}=R^{2}(Z)a^{0}_{ij}$. We then calculate from the field
equations that the following condition must
hold:
\begin{equation}
\int\frac{\partial^{n}f}{\partial
Z^{n}}\frac{v_{i}v_{j}}{(v^{T}b^{0}v)^{1/2}}~d^{3}v\Big|_{\Sigma}=0.
\end{equation}
for all~$n\geq 1$ (it is not clear whether (117) this is sufficient to
force isotropy of $f^)$).\\

From the Vlasov equation (57) we get for ~$n=1,~2$~that
conditions (117) are respectively:
\begin{equation}\chi_{ijk}a^{k}=0\end{equation}
\begin{equation}10\chi_{ijkl}a^{k}a^{l}-2a_{i}a_{j}-(a_{k}a^{k})a^{0}_{ij}=0
\end{equation}
Suppose now that we pick an~$f^{0}$~so that (118) and (119) hold, but
(117) fails to hold for some~$n>2$ (note that (118) and (119), together with
(66) and (67), amount to a set of
conditions on the first four moments of $f^0$ only, so that there are many
non-isotropic $f^0$ satisfying them, while (117) for larger and larger
$n$ imposes conditions on ever larger moments of $f^0$). A 
calculation then shows that
(114) and (115) are satisfied with~$k^{0}_{ij}=0$, so that the Weyl 
tensor vanishes at~$\Sigma$. Thus it is possible, with such an $f^0$, 
to generate a 
non-FRW Einstein-Vlasov cosmology with an initially vanishing Weyl
tensor.\\\\
\textbf{(c) Type IX}\\
Here there exists an
invariant basis for which~$C^{i}_{jk}=\epsilon_{ijk}$, and a 3-metric ~$h_{ij}$~has constant curvature
iff~$h_{ij}\propto\delta_{ij}$~in this basis.
So if we let~$a^{0}_{ij}$~be such a metric
then~$k^{0}_{ij}=0$~by (73) and the cosmology given by Theorem 6.1 is then (k=+1) FRW
with~$\tilde{f}(Z,p_{i})=f^{0}(p_{i}),~\tilde{g}_{ij}=R^{2}(Z)a^{0}_{ij}$,
for some~$R(Z)$ regardless of the choice
of~$f^{0}$. Thus FRW data in type IX is any admissible~$f^{0}$~which gives rise
to a constant curvature 3-metric.\\
As in type V one can ask whether an~$f^{0}$~can
be chosen so that the Weyl tensor vanishes at~$\Sigma$, while the
4-geometry departs from FRW shape. Here the
condition~$C_{abcd}(0)=0$~amounts to the following
\begin{equation}k^{0}_{ij}=0\end{equation}
\begin{equation}\left(9\delta_{i}^{~e}\delta_{j}^{~f}-\chi_{ij}^{~~ef}\right)S_{ef}=\frac{5}{8}Q_{ij}\end{equation}
where~$S_{ij}$~is the trace-free Ricci tensor of~$a^{0}_{ij}$~and
\begin{equation}Q_{ij}\equiv -16(b^{0})^{ln}(b^{0})^{rs}C^{t}_{ps}C^{k}_{n(i}\{\delta_{k}^{~p}\chi_{rtl|j)}+\delta_{l}^{~p}\chi_{rtk|j)}+\delta_{j)}^{~p}\chi_{rtkl}\}\end{equation} 
We have not been able to find a solution of these equations
for which~$S_{ij}\neq 0$~and it seems unlikely that there is one. If
this is so, then any~$f^{0}$~which makes the initial Weyl tensor 
vanish is FRW data.

\section*{References}
\begin{description}
\item K. Anguige and K. P. Tod 1998, ~\emph{Isotropic Cosmological
Singularities I}
\item J. Ehlers 1971,~in~\emph{General Relativity and cosmology},
ed. R. K. Sachs,~\emph{Varenna Summer School XLVII}~(New York:
Acad. Press)
\item J. Milnor 1963,~\emph{Morse Theory}~(Princeton:PUP)
\item R. Racke 1992,~\emph{Lectures on nonlinear evolution equations, Aspects of Mathematics}~vol. E19 (Vieweg)
\item A. D. Rendall, 1992,~\emph{Approaches to numerical relativity},
ed. R. d'Inverno (Cambridge: CUP)
\item A. D. Rendall 1994,~\emph{Ann. Phys.}~233,  82-96
\item A. D. Rendall 1997,~\emph{An introduction to the Einstein-Vlasov system}, in~\emph{Mathematics of Gravitation}, ed. P. Chrusciel (Banach Center Publications, Warszawa) vol. 41, part 1
\item A. D. Rendall and B. G. Schmidt 1991,~\emph{Class. Quant. Grav}
8, 985-1000
\item M. E. Taylor 1996,~\emph{Partial Differential Equations I,
Applied Mathematical Sciences}~vol. 115 (Berlin: Springer)
\item R. M. Wald 1984,~\emph{General Relativity}~(University of Chicago Press)
\end{description}
\end{document}